\pdfoutput=1

\documentclass[aps,prl,preprint,nopacs,amsmath,superscriptaddress]{revtex4}

\usepackage{epsf}
\usepackage{graphicx}
\usepackage{sidecap}
\usepackage{amsmath}

\usepackage{dcolumn}   
\usepackage{bm}        
\usepackage{amssymb}   

\begin{document}















\title{{Observation of the spin-polarized surface state in a noncentrosymmetric superconductor BiPd}}



\author{Madhab~Neupane*}
\affiliation {Department of Physics, University of Central Florida, Orlando, Florida 32816, USA}

\author{Nasser~Alidoust*}\affiliation {Joseph Henry Laboratory and Department of Physics, Princeton University, Princeton, New Jersey 08544, USA}

\author{M.~Mofazzel~Hosen*}\affiliation {Department of Physics, University of Central Florida, Orlando, Florida 32816, USA}

\author{Jian-Xin~Zhu}
\affiliation {Theoretical Division and Center for Integrated Nanotechnologies, Los Alamos National Laboratory, Los Alamos, NM 87545, USA}

\author{Klauss~Dimitri}\affiliation {Department of Physics, University of Central Florida, Orlando, Florida 32816, USA}

\author{Su-Yang Xu}
\affiliation {Joseph Henry Laboratory and Department of Physics, Princeton University, Princeton, New Jersey 08544, USA}

\author{Nagendra Dhakal}
\affiliation {Department of Physics, University of Central Florida, Orlando, Florida 32816, USA}

\author{Raman Sankar} \affiliation{Center for Condensed Matter Sciences, National Taiwan University, Taipei 10617, Taiwan}



\author{Ilya~Belopolski}
\affiliation {Joseph Henry Laboratory and Department of Physics, Princeton University, Princeton, New Jersey 08544, USA}

\author{Daniel S.~Sanchez}
\affiliation {Joseph Henry Laboratory and Department of Physics, Princeton University, Princeton, New Jersey 08544, USA}

\author{Tay-Rong Chang}
\affiliation{Department of Physics, National Tsing Hua University, Hsinchu 30013, Taiwan}

\author{Horng-Tay Jeng}
\affiliation{Department of Physics, National Tsing Hua University, Hsinchu 30013, Taiwan}
\affiliation{Institute of Physics, Academia Sinica, Taipei 11529, Taiwan}

\author{Koji Miyamoto}
\affiliation{Hiroshima Synchrotron Radiation Center, Hiroshima University, 2-313 Kagamiyama, Higashi-Hiroshima 739-0046, Japan}

\author{Taichi Okuda}
\affiliation{Hiroshima Synchrotron Radiation Center, Hiroshima University, 2-313 Kagamiyama, Higashi-Hiroshima 739-0046, Japan}

\author{Hsin~Lin}
\affiliation{Centre for Advanced 2D Materials and Graphene Research Centre,
National University of Singapore, Singapore 117546}
\affiliation{Department of Physics, National University of Singapore,
Singapore 117542}


\author{Arun~Bansil}
\affiliation {Department of Physics, Northeastern University,
Boston, Massachusetts 02115, USA}


\author{Dariusz Kaczorowski}
\affiliation {Institute of Low Temperature and Structure Research, Polish Academy of Sciences,
50-950 Wroclaw, Poland}

\author{Fangcheng Chou} \affiliation{Center for Condensed Matter Sciences, National Taiwan University, Taipei 10617, Taiwan}

\author{M.~Zahid~Hasan}
\affiliation {Joseph Henry Laboratory and Department of Physics,
Princeton University, Princeton, New Jersey 08544, USA}

\author{Tomasz~Durakiewicz}
\affiliation {Condensed Matter and Magnet Science Group, Los Alamos National Laboratory, Los Alamos, NM 87545, USA}
\affiliation {Institute of Physics, Maria Curie - Sklodowska University, 20-031 Lublin, Poland}

\date{18 June, 2013}
\begin{abstract}

\textbf{Recently, noncentrosymmetric superconductor BiPd has attracted considerable research
interest due to the possibility of hosting topological superconductivity.
Here we report a systematic high-resolution angle-resolved photoemission spectroscopy (ARPES) and spin-resolved ARPES study of the normal state electronic and spin properties of BiPd. Our experimental results show the presence of  a surface state at higher-binding energy with the location of Dirac point at around 700 meV below the Fermi level. The detailed photon energy, temperature-dependent and spin-resolved ARPES measurements complemented by our first-principles calculations demonstrate the existence of the spin polarized surface states at high-binding energy. 
The absence of such spin-polarized surface states near the Fermi level negates the possibility of a topological superconducting behavior on the surface.
Our direct experimental observation of spin-polarized surface states in BiPd
provides critical information that will guide the future search for topological superconductivity in noncentrosymmetric materials.}

\end{abstract}
\date{\today}
\maketitle


Recently, noncentrosymmetric (NCS) superconductors (SCs) have attracted considerable
research interest due to the possibility that they host several exotic states \cite{note, Bauer, Sigrist}. The
lack of inversion symmetry creates an asymmetric potential gradient, which may
split the electron bands by lifting the spin degeneracy, allowing hybrid-pairing of
spin-singlet and spin-triplet states within the same orbital channel \cite{Sigrist, PRB_2011, PRB_2014, arXiv_2014}. Furthermore, with the advent of topological insulators \cite{Hasan, Hsieh, SCZhang, Hasan_review_2, Neupane, Superconducting_TI, Neupane_2, Neupane_3, Neupane_4, Xia, Hasan_review_3, Bansil_review}, it was
recently proposed that NCS SCs with a strong spin-orbit coupling are potential candidates
for realizing topological superconductivity \cite{Heusler_2010, CKLu, Hsin_Heusler,Ludwig_classification}, holding promise for hosting protected Majorana surface states \cite{Majorana, Superconducting_TI, Vishwanath_PRL2011}.
But so far, evidence of topological surface states in noncentrosymmetric materials is still lacking. Such convincing evidence may be obtained by angle- and spin-resolved photoemission, which provides an energy- and momentum-resolved probe of the electronic and spin structure.


 A NCS system provides a good platform to look for exotic states such as 
the Weyl semimetal in TaAs \cite{Suyang_Science} and the topological nodal-line phase in PbTaSe$_2$ \cite{Guang}. 
Another NCS SC BiPd can provide a platform to study the interplay
of spin-orbit coupling (SOC) effects with superconductivity. This compound undergoes a
structural transition from $\beta$ - BiPd (orthorhombic) to $\alpha$ - BiPd (monoclinic) at 210 $^{\circ}$C and
then becomes superconducting at the transition temperature ($T_c$) $\sim$ 3.7 K \cite{PRB_2011, PRB_2014}. In comparison to many other NCS
SCs, BiPd is a weakly correlated compound possessing a heavy atom, Bi. Measurements
by point-contact spectroscopy \cite{tunelling} and nuclear quadrupole resonance (NQR) \cite{NMR} indicate a
complex gap structure in BiPd, which might be caused by the lack of inversion symmetry
\cite{PRB_2011, PRB_2014}. Recent scanning tunneling microscopy (STM) measurements reveal that the
superconducting state of BiPd appears to be topologically trivial, consistent with
Bardeen-Cooper-Schrieffer theory with an $s$-wave order parameter \cite{arXiv_2014}. 
 Further momentum and spin-resolved experimental evidence is highly
desirable in order to establish the presence of topological Dirac surface states, and to study the underlying microscopic mechanism of superconductivity.
A detailed high-resolution angle-resolved photoemission spectroscopy (ARPES) and spin-resolved ARPES study is necessary to prove or disprove the topological nature of the surface states in the normal state of BiPd.
Such a characterization of the normal state is currently missing, but necessary and needed as a first step towards opening a discussion on the relationship between superconductivity and
the topological properties of NCS systems.

In this paper, we report the experimental observation of spin-polarized surface states in the NCS material  BiPd using ARPES and spin-resolved ARPES. Our experimental results show the presence of  a surface state at higher-binding energy, with the Dirac node located at around 700 meV below the Fermi level. 
Furthermore, the Dirac node located at around 650 meV below the Fermi level is also observed for different surface cleaving, providing direct evidence for the noncentrosymmetic crystal structure presented in BiPd. 
Our measurements suggest that the observed states may be topological in nature. Our results are further supported by our first-principles calculations. The absence of spin-polarized surface states near the Fermi level of BiPd negates the possibility that this system might host topological superconductivity on the surface.


\bigskip
\bigskip
\textbf{Results}

\textbf{Crystal structure and sample characterization.}
The crystal structure of BiPd at low temperatures ($< $ 210 $^{\circ}$C) has a monoclinic unit cell with $a= 5.63$ $\AA$,  $b= 10.66$ $\AA$, $c= 5.68$ $\AA$, $\alpha = \gamma =90^{\circ}$, and $\beta =101^{\circ}$ with the $b$ axis being as its unique axis (see Fig. 1a; Ref. \cite{PRB_2011}).
Detailed characterization of the single crystals used in our study indicated their high quality (see Supplementary Figure 1-7; Supplementary Note 1-4; Supplementary References). They exhibit a simple metallic behavior in the normal state and a sharp superconducting transition at  $T_c$ $\sim$ 3.7 K. The low-temperature magnetic susceptibility data is shown in Fig. 1b (see also Supplementary Figure 1 and Supplementary Note 1 for additional data and discussion).
A schematic bulk Brillouin zone is shown in Fig. 1c, where the projected surface along (010) is also illustrated. 
In order to experimentally identify its electronic structure, we study the electronic
structure of BiPd on the cleaved (010) surface. Figure 1d  (a picture of the single crystal measured is displayed in the inset of Fig. 1d) shows
momentum-integrated ARPES spectral intensity over a wide
energy window. Sharp ARPES intensity peaks at binding energies
$E_B$ $\sim$  23 and 26 eV corresponding to the bismuth 5$d_{3/2}$ and
 5$d_{5/2}$ energy levels are observed.

 \bigskip
\bigskip

 \textbf{Electronic structure of BiPd.}
 We study the overall electronic structure of BiPd using ARPES. Figure 2a shows an ARPES dispersion map in a 1.3 eV binding energy window, where several dispersive bands within the valence band are identified. 
Moreover, several crossing or metallic bands in the vicinity of  the Fermi level are observed.  Remarkably, a nearly linearly dispersive Dirac cone-like state is observed at the Brillouin zone center, showing a Dirac node located at a binding energy of $E_B$ $\sim$ 700 meV. The Dirac-like state can be observed in the region of the blue rectangle in the Fig. 2a, at the binding energy region of 500$-$900 meV.  At the Fermi level only the metallic bands, but no other Dirac like linearly dispersive bands are observed. On the other hand, the linearly dispersive Dirac-like bands are found to be in the region of higher-binding energies.

Figure 2b-e shows the Fermi surface and constant energy contour plots, respectively. In the vicinity of the Fermi level, many metallic bands are observed (Fig. 2b). 
We also study the ARPES measured constant energy contour maps
(Fig. 2c-e). At the Fermi level, the constant energy contour
consists of many metallic pockets. With increasing
binding energy at $\sim$ 600 meV, the circular pocket formed by the upper Dirac cone is observed (Fig. 2c). On increasing the binding energy, the size of the pocket decreases and eventually shrinks to a point (the Dirac point) near $E_B$ $\sim$ 700 meV (Fig. 2d). At even higher binding energies, the nearly circular pocket formed by the lower Dirac cone is observed (for $E_B$ $\sim$ 750 meV in Fig. 2e).

To reveal the nature of the states observed in BiPd, we performed photon energy and temperature-dependent ARPES, as well as spin-resolved ARPES measurements which are further complemented by calculations.
Figure 3 shows the energy-momentum cuts measured with varying photon energies from 30 to 58 eV with a 4 eV energy step (see Supplementary Figure 2-4 and  Supplementary Note 2 for additional ARPES data and discussion). Clear $E$-$k$ dispersion of the bulk bands is observed. Remarkably, the dispersion of the linearly dispersive states at high-binding energy (500-900 meV) is found to be unchanged with respect to the varying photon energy, supporting the two-dimensional nature of this state.
It is important to note that the Dirac-like two-dimensional states are not found to be perfectly linear in energy-momentum axis. Furthermore, it is important to recall that in real materials such as pure Bi, graphene, or topological insulators, the Dirac cones are never perfectly linear over a large energy window yet they can be approximated as linear within a narrow energy window around the Dirac point. 
This linear part represents the massless dispersion, in contrast to the large effective mass of conventional band electrons in other materials.
Moreover,  we note that the weak asymmetry of the surface bands with respect to k and -k is probably coming from the matrix element effects, and from the fact that these measured spectra were slightly off from the ${\bar{X}}-{\bar{\Gamma}}-{\bar{X}}$ direction.


To test the robustness of the surface state observed in BiPd, we have performed a temperature-dependent measurement as shown in Fig. 4a.  On raising the temperature, the Dirac-like surface states survive even at room temperature, which establishes that  the Dirac-like surface states are robust to the rise of temperature and potential surface contamination, resulting from increased partial pressure of residual gases released upon heating,
see Fig. 4a. Furthermore, the observed Dirac-like states are found to be robust against thermal cycling (20 K - 300 K - 20 K), since lowering the temperature back down to 20 K results in the similar
spectra with the strong presence of Dirac-like state features (see the panel of Fig. 4a with the note of Re\_20K).

 \bigskip
\bigskip
 \textbf{NCS electronic signature.}
To better understand the electronic structure observed with ARPES, we perform first-principles
calculations of the bulk band structure (Supplementary Figure 5; Supplementary Note 3) and slab calculations of BiPd using the generalized gradient approximation plus spin-orbit coupling (SOC) method (see Fig. 4b,c for the calculated electronic structure for the top and bottom surfaces, respectively). We note that the band structure calculations were performed on a conventional unit cell for the bulk system and on  a supercell for the slab structure, which leads to folding of the states at the S point to the $\Gamma$ point. Our slab calculations show that the surface states are predicted to be at the $\Gamma$ point with the location of the Dirac nodes at $\sim$ 0.5 and 0.6 eV below the Fermi level, one of which comes from the top surface and the other from the bottom surface. The Dirac node of the surface state coming from the top and bottom surfaces are located at different binding energies (not degenerate), which comes from the fact that BiPd lacks inversion symmetry.

Since the calculations were made for slab geometry, the obtained band structure always involves bands from two surfaces, one on the top and the other on the bottom of the slab. Experimentally, one can only measure one surface (corresponding to either top or bottom surface of the slab geometry) at a time.  In principle, it should be possible to measure either the top or bottom surface on the two sample pieces based on the cleaving.  Therefore, depending on whether the cleaved surface is a top or bottom one, the location of the Dirac point energy is predicted to be different. 
Experimentally, the location of the Dirac point energy is found to be different based on the cleaving either the top or the bottom surface, in agreement with our calculations.  
Actually, we experimentally observed the different Dirac point energy for top and bottom surface cleaving. For the top surface the Dirac point is $\sim$ 700 meV, while for the bottom surface cleaving Dirac point is $\sim$ 650 meV from the Fermi level (Supplementary Figure 4; Supplementary Note 3). We note that such two terminations of the crystal surface in BiPd are directly observed by STM topographic image \cite{arXiv_2014}.

To further understand the nature of the observed Dirac-like band, we measured the in-plane spin polarization or spin-texture properties of BiPd.
Spin-resolved ARPES measurements were performed on the top-cleaved surface of BiPd. Two spin-resolved energy-dispersive curves (EDCs) are shown at momenta of $\sim$  $\pm$ 0.1 on the opposite sides of the Dirac-like dispersion. The blue dash lines on the ARPES dispersion map shown in Fig. 5 mark the approximate momentum positions for the spin-resolved energy distribution curves. The spin-resolved EDCs and corresponding net spin polarization are shown in Fig. 5b,c for momentum positions \#1 and \#2. The obtained spin data presented in Fig. 5b,c show observable net spin polarization. Importantly, the observed spin polarization at one branch of the cone is opposite with that of another cone, thus confirming the spin-momentum locking behavior. Spin-resolved ARPES data reveals a characteristic helical spin-texture, suggesting the possible topological origin of the states we observed. However, we caution that two-band crossings at a time-reversal invariant momentum should have an opposite value of the spin polarization independent of any topological invariant.

\bigskip

\bigskip
\textbf{Discussion}

The experimental realization of the topological insulator phase in a NCS crystal structure is an object of intense  research. Such a system may be utilized in testing several proposed exotic phenomena, such as crystalline-surface-dependent topological electronic states, pyroelectricity, and natural topological $p$-$n$ junctions \cite{theory}. 
Recently, the first principles calculations predicted III-Bi
to be an inversion asymmetric topological insulator with large band gap possessing intrinsic topologically protected edge states and forming quantum spin Hall systems \cite{theory}, but these have not yet been realized experimentally. At the same time, 
the proposal of topological insulating nature in an inversion asymmetric compound BiTeCl still remains under debate \cite{YLChen_BiTeCl, Hugo_BiTeCl}. 
Furthermore, the small bulk band gaps of the realized inversion asymmetric topological insulators severely limit the manipulation and control of the topological surface states.

We note that BiPd consists of many bands near the Dirac-like states, which makes this system very complicated with no global band gap in the normal state. The topology in BiPd is not well defined like in Bi$_2$Se$_3$. 
In ARPES experiments, the samples were cleaved and measured in an ultrahigh vacuum environment that kept its surface clean during the measurements. The surface cleaving generates surface potential due to the surface charges, which may develop a large effective pressure along the $b$ (perpendicular to the surface, see Fig. 1a) direction and thus drive the crystal (possibly the several top layers) into the topological insulator phase. Our calculations on strained BiPd show the bulk band gap opening, (Supplementary Figure 7; Supplementary Note 4). This study will enable further discussion on the theoretical origin of the topological order in NCS materials.




BiPd is a superconductor below $T_c$ $\sim$ 3.7 K. Since the spin polarized surface state is located at high binding energy, it negates the topological superconductivity behavior on the surface at its native Fermi level. 
However, by electrical gating or surface deposition, the Fermi level can be tuned near the Dirac surface state, which provides an  opportunity  to  realize the topological superconductivity in this noncentrosymmetic material.  
\newline
\newline
\textbf{Methods}
\newline
\textbf{Crystal growth and characterization.}
Single crystals of BiPd were grown by a modified Bridgman method as described elsewhere \cite{PRB_2011}. 
The crystals were characterized by means of X-ray diffraction, energy dispersive X-ray spectroscopy, magnetic susceptibility, electrical resistivity and heat capacity measurements, using standard commercial equipment.   
\newline
\newline
\textbf{Electronic structure measurements.}
Synchrotron-based ARPES measurements of the electronic structure were performed at the Advanced Light Source (ALS), Berkeley at Beamline 10.0.1 and Stanford Synchrotron Radiation Lightsource (SSRL) at Beamline 5-4 both equipped with a high efficiency R4000 electron analyzer. The energy resolution was set to be better than 20 meV for the measurements with the synchrotron beamline. The angular resolution was set to be better than  0.2$^{\circ}$ for all synchrotron measurements. 
Samples were cleaved in situ and measured at 10 - 80 K in a vacuum better than 10$^{-10}$ torr. They were found to be very stable and without degradation for the typical measurement
period of 20 hours.
\newline
\newline
\textbf{Spin-resolved ARPES measurements.}
Spin-resolved ARPES measurements were performed at the ESPRESSO end station installed at Beamline-9B of the Hiroshima Synchrotron Radiation Center (HiSOR), Hiroshima, Japan, equipped with state-of-the-art very low-energy electron diffraction (VLEED) spin detectors utilizing preoxidized Fe(001)-p(1 x 1)-O targets \cite{Spin, Spin_1}. The two spin detectors are placed at an angle of 90 $^{\circ}$ and are directly attached to a VG-Scienta R4000 hemispheric analyzer, enabling simultaneous spin resolved ARPES measurements for all three spin components. We also measured the x and z component of the spin which shows negligible spin polarization. 
\newline
\newline
\textbf{First-principles calculations.}
The first-principles calculations were based on the generalized gradient approximation (GGA) \cite{GGA} using the projector augmented-wave method \cite{PAW} as implemented in the VASP package \cite{VASP,VASP_1}. The experimental crystallographic structure was used \cite{expt} for the calculations. The spin-orbit coupling was included self-consistently in the electronic structure calculations with a 6 $\times$ 4 $\times$ 5 Monkhorst-Pack $k$-mesh.
In order to simulate surface effects, we used 1 $\times$ 5 $\times$ 1 supercell for the (010) surface, with a vacuum thickness larger than 20 \AA.
\newline
\newline
\textbf{Data availability}. All relevant data are available from the corresponding author upon request.

\bigskip
\bigskip
\bigskip
\hspace{0.5cm}
\textbf{Acknowledgements}
\newline

M.N. is supported by the start-up fund from University of Central Florida and LANL LDRD Program.  
T.D. was supported by NSF IR/D program. D.K. was supported by the National Science Centre (Poland) under research grant 2015/18/A/ST3/00057. J.-X. Zhu is supported by the Center for Integrated Nanotechnologies, a U.S. DOE Office of Basic Energy Sciences user facility, in partnership with the LANL Institutional Computing Program for computational resources.
The work at Princeton and synchrotron x-ray-based measurements are supported by the Office of Basic Energy Sciences, US Department of Energy (DOE) Grant No. DE-FG-02-40105ER46200. The work at Northeastern University is supported by the DOE, Office of Science, Basic Energy Sciences grant number DE-FG02-07ER46352, and benefited from Northeastern University's Advanced Scientific Computation Center (ASCC) and the NERSC supercomputing center through DOE grant number DE-AC02-05CH11231.
H.L. acknowledges the Singapore National Research Foundation for the support under NRF Award No. NRF-NRFF2013-03. 
T.R.C. and H.T.J. were supported by the National Science Council, Taiwan. We also thank NCHC, CINC-NTU, and NCTS, Taiwan for technical support.
The measurements at HiSOR were performed with the approval of the Proposal Assessing Committee of HSRC (Proposal No. 15-A-66).
We thank Sung-Kwan Mo and Makoto Hashimoto for beamline assistance at the LBNL and the SSRL.

 \bigskip
\hspace{0.5cm}
\textbf{Authors Contributions}
\newline
M.N. designed the study; D.K., R.S. and F.-C.C. synthesized the samples and performed the transport characterization; M.N. performed the spectroscopy measurements with the help of  N.A., S.-Y.X., I.B., D.S., M.M.H., K.D., N.D., K.M., T.O., M.Z.H., T.D.; J.X.Z., T.-R.C., H.-T.J, A.B. performed the calculations; M.N. wrote the manuscript and discussed with all the co-authors. 
M.N. was responsible for the overall research direction, planning and integration among different research units.

 \bigskip
\hspace{0.5cm}
\textbf{Additional Information}
\newline
\textbf{Competing financial interests}: The authors declare no competing financial interests.
\newline
\textbf{Supplementary Information} accompanies this paper at http://www.nature.com/
naturecommunications.
\bigskip

\*Correspondence and requests for materials should be addressed to M.N. (Email: Madhab.Neupane@ucf.edu).

\begin{figure*}
\centering
\includegraphics[width=14.5cm]{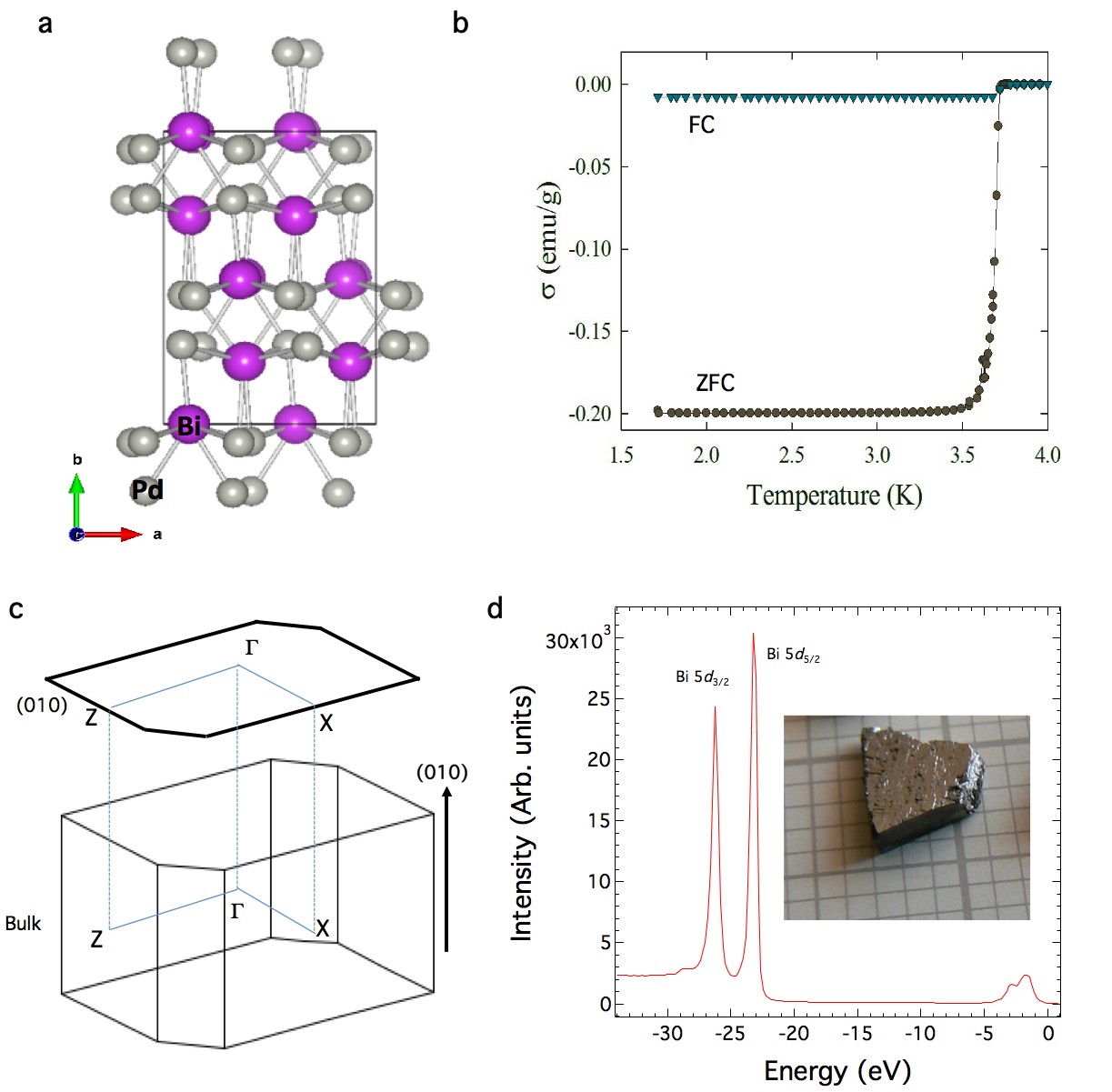}
\caption{\textbf{Crystal structure and sample characterization of BiPd}. 
\textbf{a,} Crystal structure of BiPd with the $b$ axis shown as its unique axis. It crystallizes in a monoclinic structure at low temperatures.  \textbf{b,} The magnetic susceptibility as a function of temperature showing a sharp superconducting transition temperature at $\sim$ 3.7 K in the field of 20 Gs. \textbf{c,} Schematic drawing of surface and bulk Brillouin zones, corresponding to a primitive cell,  are shown. We note that all electronic structure calculations were performed on a conventional unit cell for the bulk and a supercell for the slab. High-symmetric points are also marked.
\textbf{d,} Core level spectroscopic measurement of BiPd showing sharp peaks of Bi 5$d$. The inset shows a photograph of the BiPd sample.}
\end{figure*}

\begin{figure*}
\centering
\includegraphics[width=16.5cm]{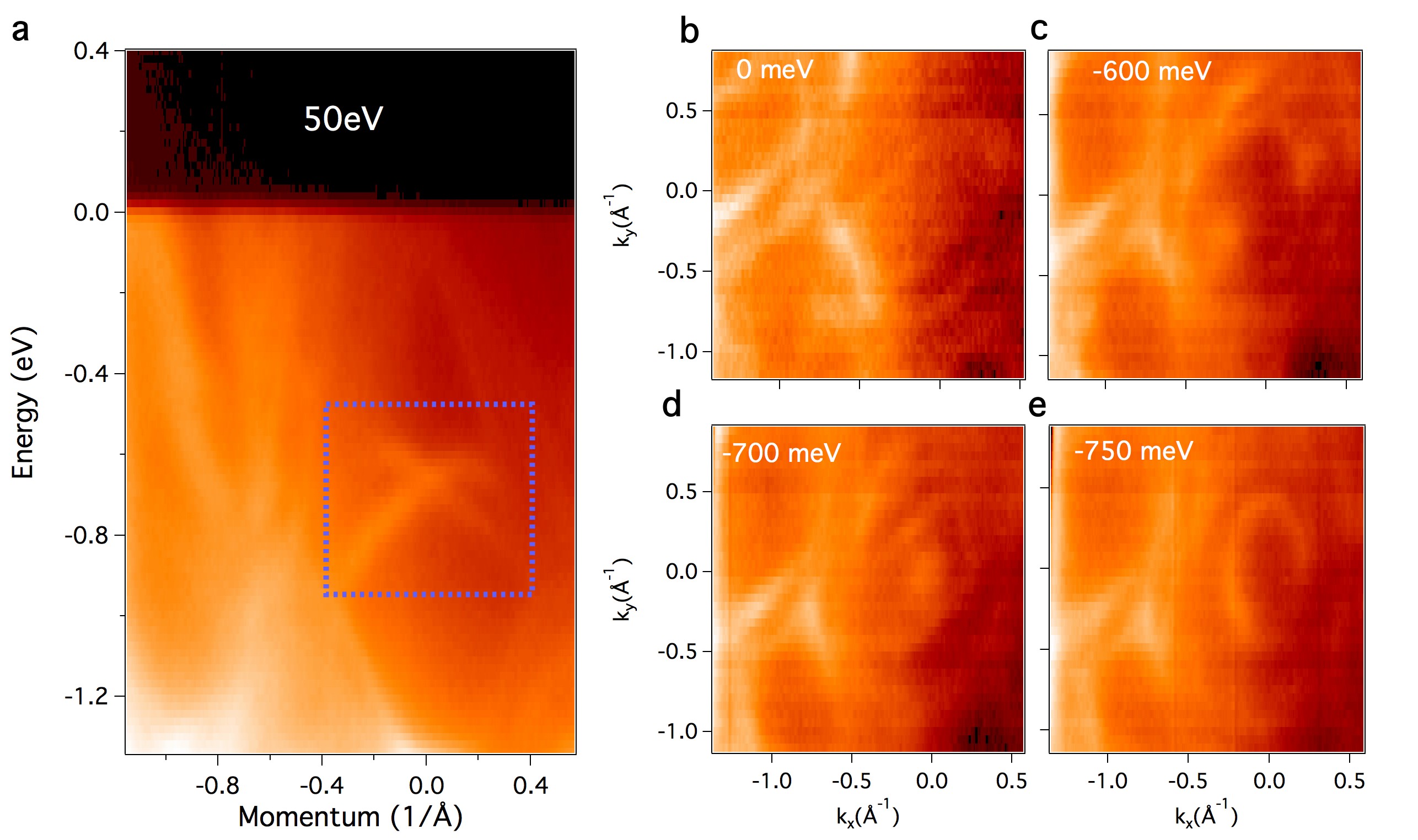}
\caption{\textbf{Electronic structure of BiPd.} \textbf{a,} Dispersion map of BiPd  along the zone centre obtained by using incident photon energy of 50 eV at a temperature of 10 K. The blue rectangle in the binding energy range of about  500 meV  to 900 meV shows the linearly dispersive states. \textbf{b,} Fermi surface map. \textbf{c,} Constant energy contour at binding energy of 600 meV shows the intensity map in the region above the Dirac point. \textbf{d-e,} The constant energy contours at 700 meV and 750 meV show the intensity map at around and below the Dirac point, respectively. The values of binding energies are noted on the plots. Data were collected at ALS BL 10.0.1.}
\end{figure*}

\begin{figure*}
\centering
\includegraphics[width=16.5cm]{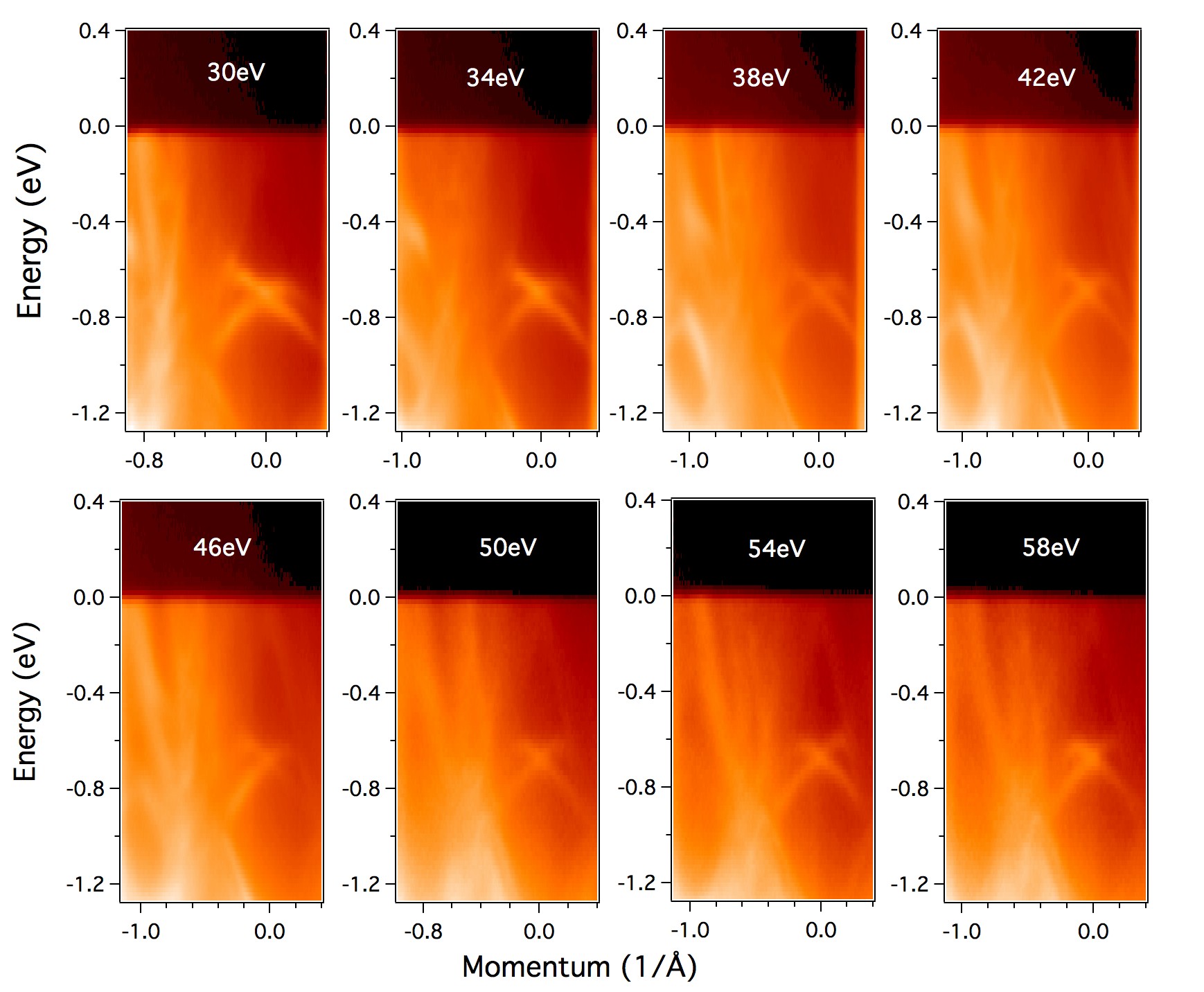}
\caption{\textbf{Photon energy dependent ARPES dispersion maps.}
The measured photon energies are noted on the plots. The linearly dispersive states at a binding energy of around 700 meV do not show any dispersion with photon energy, which indicate its two dimensional nature. These data were collected along ${\bar{\Gamma}}-{\bar{X}}$ high symmetry direction at ALS BL 10.0.1 at a temperature of 10 K. Additional photon energy data are presented in Supplementary Information.}
\end{figure*}



\begin{figure*}
\centering
\includegraphics[width=17.5cm]{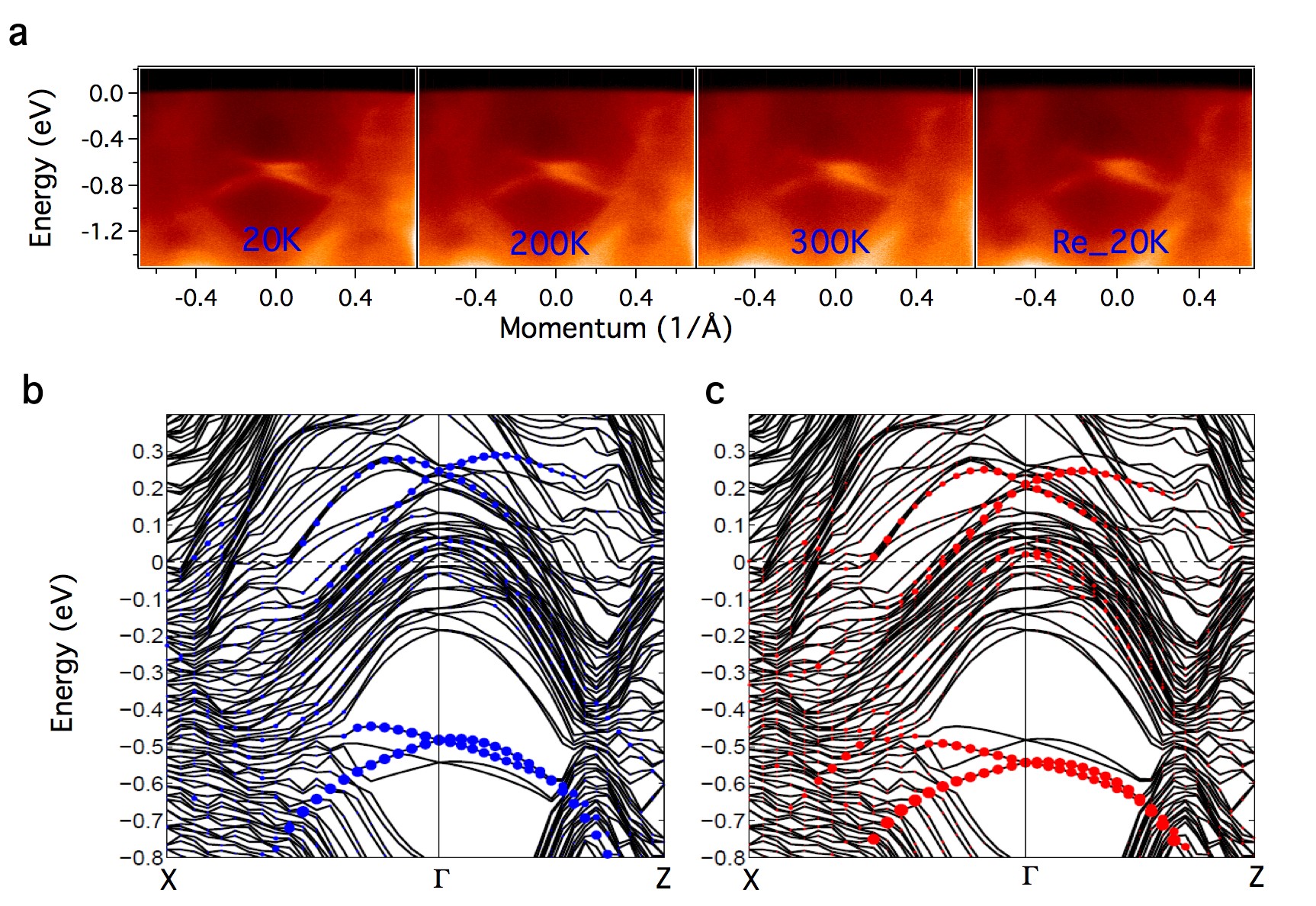}
\caption{\textbf{Temperature dependent spectra and first-principles calculations.} \textbf{a,} ARPES energy momentum dispersion maps measured using a photon energy of 30 eV along the ${\bar{X}}-{\bar{\Gamma}}-{\bar{X}}$ momentum space cut-direction with varying temperatures. The measured values of temperature are noted on the plots. The panel with the note of Re\_20K is the spectrum measured after thermal cycling (20 K$\rightarrow$300 K$\rightarrow$20 K). These data were collected at SSRL BL5-4 with a photon energy of 30 eV. \textbf{b,} Slab calculations of BiPd for the top surface and \textbf{c,} the bottom surface along the high symmetry lines. The blue (red) dots represent surface states for the top (bottom) surface. Details of the calculation method is given in Method Section and additional calculation plots are shown in Supplementary Information.}
\end{figure*}

\begin{figure*}
\centering
\includegraphics[width=17cm]{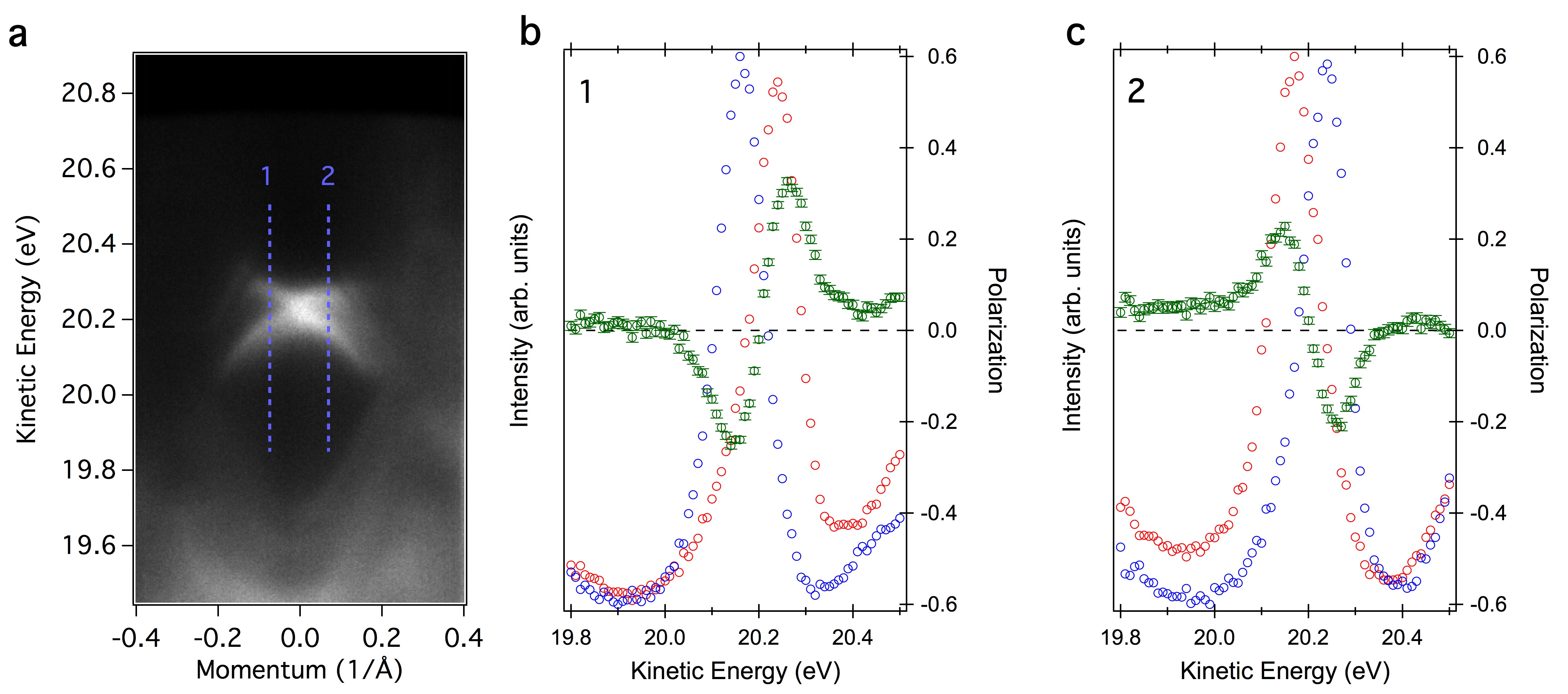}
\caption{\textbf{Spin-resolved ARPES results.} \textbf{a,} Spin-integrated ARPES spectrum of surface states. The blue dashed lines and number indicate the
momentum positions at which SR-ARPES measurements were carried out. \textbf{b,} Spin-resolved EDCs  (blue and red circles for up and down spin) and the in-plane spin polarization (red circle with error bar -right axis) at momentum position \#1. \textbf{c,} Same as \textbf{b,}  for momentum position \#2. Helical spin-texture is observed in BiPd. Error bars represent the experimental uncertainties (standard deviation) in determining the spin polarization. 
Spin-resolved measurements were carried out with photon energy of 25 eV and temperature of 20 K at Beamline-9B of the Hiroshima Synchrotron Radiation Center (HiSOR), Hiroshima, Japan.}
\end{figure*}

\newpage

\setcounter{figure}{0}

\renewcommand{\figurename}{\textbf{Supplementary figure}}

\clearpage

\textbf{
\begin{center}
{\Large \underline{Supplementary Information}: \\
Observation of the spin-polarized surface state in a noncentrosymmetric superconductor BiPd}
\end{center}
}

\vspace{0.35cm}

\textbf{This files includes:}

\textbf{
\begin{tabular}{l l}
  Supplementary Figures \\ 
   Supplementary Notes \\
 Supplementary References \\
\end{tabular}}

\clearpage










\begin{figure*}[h!]
\includegraphics[width=16.0cm]{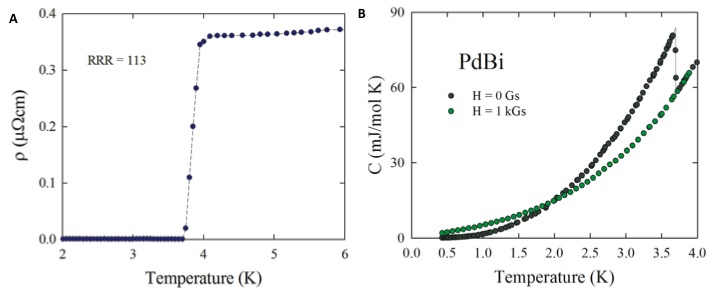}
\caption{{Sample characterization of BiPd.  (a) Temperature dependence of electrical resistivity within (010) plane showing a sharp transition at $T_c$ $\sim$ 3.7 K.  (b) The temperature dependence of the specific heat capacity with no field (black filled circles) and with field (red filled circles).}}
\end{figure*}

\clearpage

\begin{figure*}[h!]
\includegraphics[width=16.50cm]{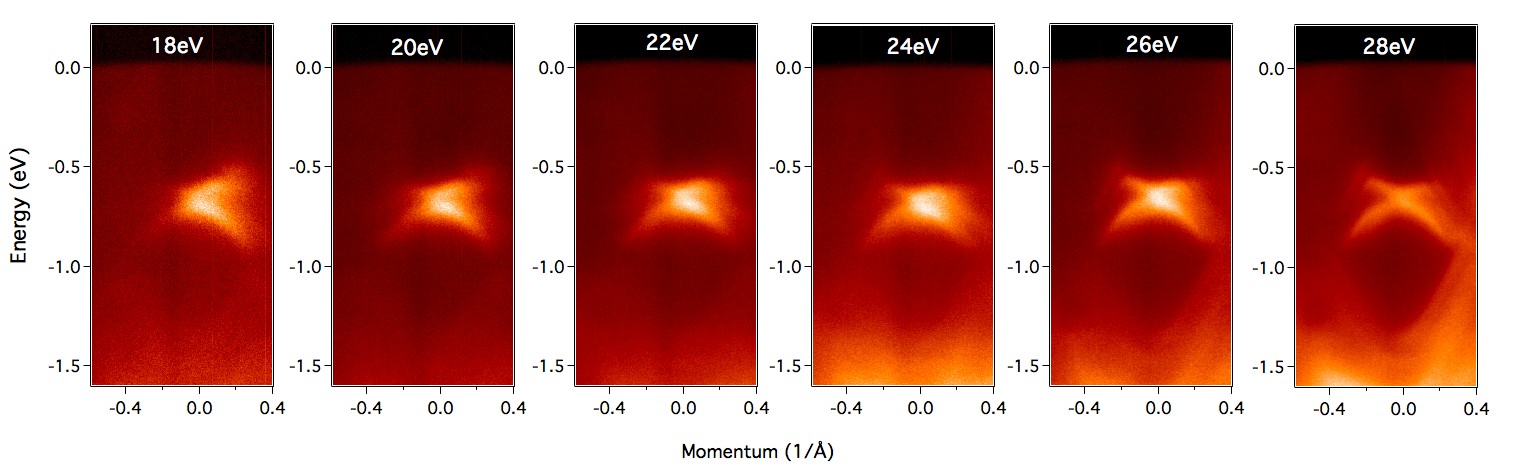}
\caption{{ Photon energy dependent spectra. ARPES dispersion maps measured at zone center $\Gamma$ point with varying photon energy. The measured photon energy is noted on the spectra. 
These data were collected at SSRL beamline 5-4 at a temperature of 10 K. }}
\end{figure*}

\begin{figure*}[htbp]
\centering
\includegraphics[width=17cm]{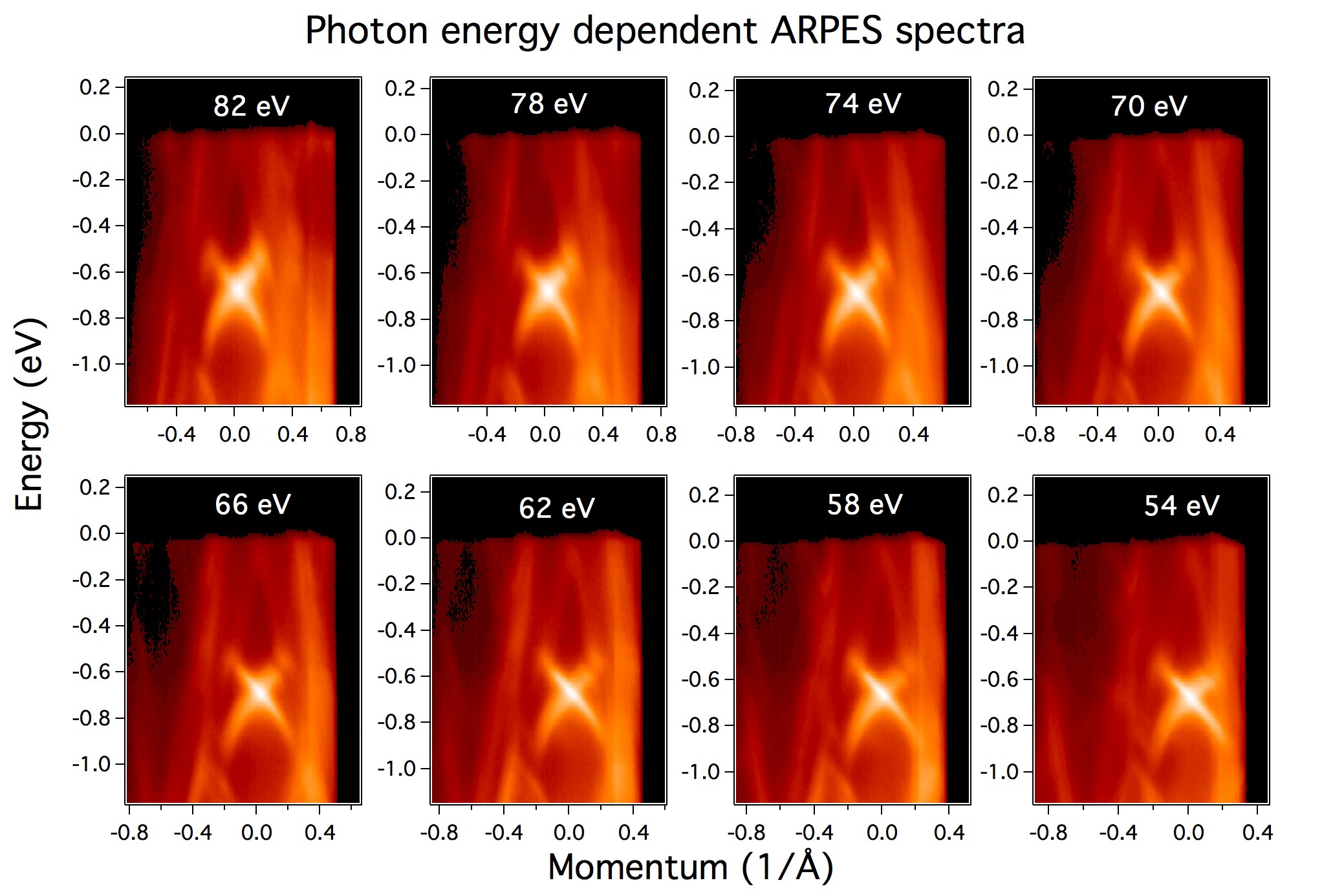}
\caption{Photon energy dependent ARPES measurements:  Measured photon energies are marked on the plots.}
\end{figure*}

\clearpage

\begin{figure*}[h]
\centering
\includegraphics[width=17cm]{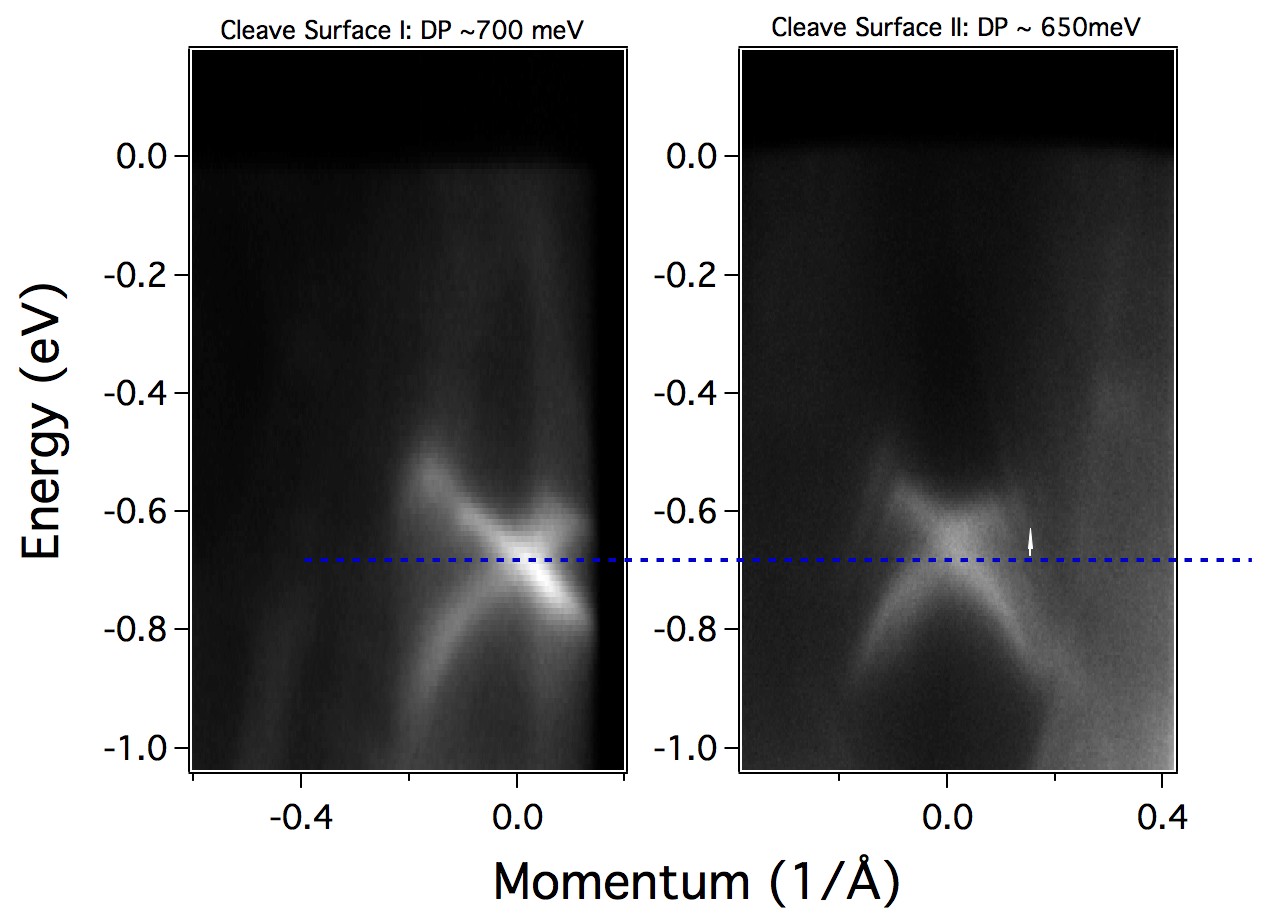}
\caption{ARPES energy momentum spectra obtained from two different cleaved surfaces. Blue dash line shows the energy position of the Dirac point for the cleaved surface I (left) and a small arrow shows the slightly different energy position for the cleaved surface II (right). These two spectra were measured in similar experimental conditions.}
\end{figure*}

\clearpage

\begin{figure*}[h!]
\includegraphics[width=16.5cm]{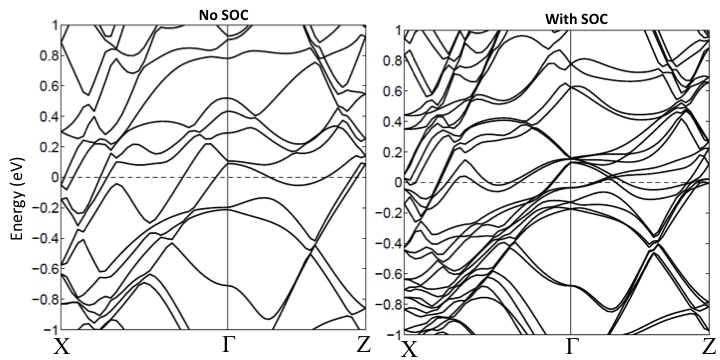}
\caption{{Calculation results.}  Calculated bulk band structure with the projection along the (010) direction considering no spin-orbit coupling (left) and with spin-orbit coupling (right).}
\end{figure*}

\clearpage

\begin{figure*}[h!]
\centering
\includegraphics[width=17cm]{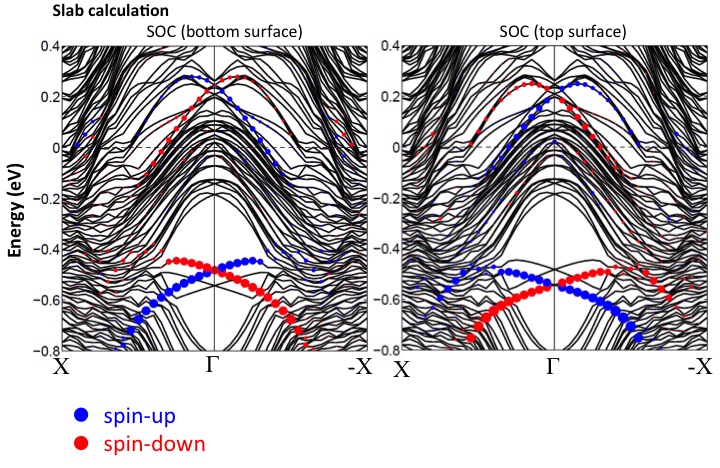}
\caption{In-plane spin polarization. Slab calculations at around the $\Gamma$ point showing the helical in-plane spin structure of the surface band for the top surface (left) and bottom surface (right).}
\end{figure*}

\begin{figure*}[h!]
\centering
\includegraphics[width=17cm]{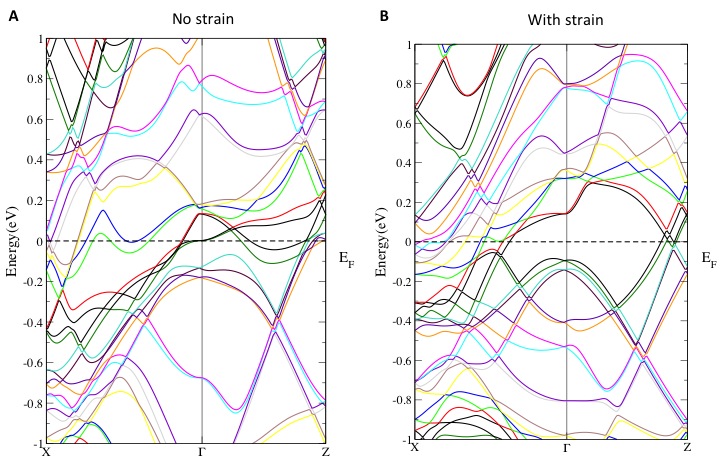}
\caption{ (a) Band structure of BiPd without considering strain. (b) Band structure of BiPd with applying strain. We apply strain with $b'$=0.81$b$, $a'$=$a$/0.9, and $c'$=$c$/0.9 to keep the volume unchanged.}
\end{figure*}

\clearpage


{\large {Supplementary Note 1}}
\vspace{0.1cm}

\textbf {Sample characterization}


Besides the Meissner state evidenced in the magnetic susceptibility data, the superconductivity in BiPd was proved by means of electrical transport and heat capacity measurements. The electrical resistivity was measured on a bar-shaped specimen with the dimensions 0.17 mm $\times$ 0.52 mm $\times$ 1.48 mm (the estimated error in geometrical factor is less than 10$\%$) and the electrical contacts were made of silver epoxy paste. The measurement was performed with current flowing within the (010) plane. Fig. S1a displays the low-temperature electrical resistivity with a sharp drop to zero resistance at $T_c$ $\sim$ 3.7 K. It is worth noting that just above $T_c$, the resistivity is only $\rho_{0}$ = 0.36 $\mu$$\Omega$cm, and the ratio between the resistivity at room temperature and  $\rho_{0}$ (residual resistivity ratio RRR) is as large as 113, which signals the high-quality of the single crystal studied. At the same temperature, the specific heat forms a distinct anomaly (see Fig. S1b), which corroborates the bulk nature of the superconducting state. The obtained results are in very good agreement with the literature data \cite{PRB_2011}.

\vspace{0.5cm}
{\large {Supplementary Note 2}}
\vspace{0.1cm}

\textbf{{Photon energy dependent ARPES measurements}}

We present ARPES measured dispersion maps at the $\Gamma$ point, with photon energies from 18 eV to 28 eV with every 2 eV energy step in  Fig. S2. At a low photon energy, the bulk bands are suppressed and Dirac like surface states are highly enhanced.  
Most importantly, our wide range of photon energy measurements (from 18 eV to 82 eV)  reveal a negligible dispersion of the Dirac like states confirming its two-dimensional nature as shown in Ref. \cite{Xia, Neupane}.

To further confirm the nature of the states in BiPd, we cleaved another sample. The ARPES  dispersion map of this sample is shown in Fig. S3. ARPES spectra were measured from 56 eV to 82 eV with every 4 eV energy step. These data show that there is no clear dispersion with photon energy, which suggests its two-dimensional nature.

Since the calculations were made for slab geometry, the obtained band structure always involves bands from two surfaces, one on the top and the other on the bottom of the slab. Experimentally, one can only measure one surface (corresponding to either top or bottom surface of the slab geometry) at a time.  In principle, it should be possible to measure either top or bottom surface on the two sample pieces based on the cleaving.  Therefore, depending on whether the cleaved surface is a top or bottom one, the location of the Dirac point energy is predicted to be different. 

To approach both the top and bottom surfaces, we cleaved multiple samples and measured in identical experimental conditions such as sample temperature, photon energy, and the light polarization etc. 
Experimentally, the location of the Dirac point energy is observed to be different based on the cleaving of either the top or bottom surface, which is consistent with our calculations.  We observed a different Dirac point energy for the top and bottom surfaces, based on the cleaving. For the top surface, the Dirac point is located at about 700 meV from the Fermi level, whereas the Dirac point is located at about 650 meV from the Fermi level for the bottom surface (see Fig. S4). It is a direct consequence of the centrosymmetric behavior of BiPd. We note that we cleaved multiple samples and most of cleaved surfaces show the Dirac point located at about 700 meV. We note that such two terminations of the crystal surface are directly observed by STM topographic image \cite{arXiv_2014}.

\vspace{0.5cm}
{\large {Supplementary Note 3}}
\vspace{0.1cm}

\textbf{{Calculations}}

Figures S5 and S6 show the calculated band structure along the high symmetry lines. 
In Fig. S5, the bulk band structure is shown along various high-symmetry lines without considering spin-orbit coupling (left) and with considering spin-orbit coupling (SOC) (right). These results are in agreement with Ref. \cite{arXiv_2014}. SOC leads to the spin splitting of the bands as well as the shifting the energy position of the bands. The orbital character of the bands near the Fermi level is mostly contributed from Bi 6$p$ and Pd 4$d$ and they hybridized strongly leading to the complex Fermi surface as shown in Fig. 2 of the maintext.

Furthermore, our slab calculations show a helical in-plane spin structure (see Fig. S6) of the surface bands at the $\Gamma$ point, which suggests that the surface states we experimentally observed in our ARPES measurements may be topological in nature. Spin calculation is consistent with the results obtained with spin-resolved ARPES measurements.


Our systematic experimental data and calculations reveal the following important properties of the electronic structure in BiPd:
\newline
(1) The bands located around the binding energy range of  900 meV to 500 meV are found to exhibit nearly linear (Dirac like) in-plane dispersion.
\newline
 (2) No observable out-of-plane (k$_z$) dispersion is observed for these Dirac like bands. 
 \newline
 (3) Our spin-resolved ARPES results and slab calculations show the helical  spin structure of these surface states.
 \newline
 (4) We found an odd number of spin-momentum locked constant energy contour per Brillouin zone. This implies that if the Fermi level is tuned in the vicinity of the Dirac point, the odd number of Fermi surfaces formed by spin-momentum locked states can be found.
 \newline 
 (5) The spin-momentum locked states wind around the time-reversal invariant momentum point.
 
 All these above properties suggest the possibility of the topological surface state in BiPd. However, we caution that such properties are also present in noncentrosymmetric materials with high values of spin orbit coupling.


\vspace{0.5cm}
{\large {Supplementary Note 4}}
\vspace{0.1cm}

\textbf{{Effect of strain}}

We note that in ARPES experiments, the samples were cleaved and measured in the ultrahigh vacuum environment that kept the surface clean during the measurements. The sample cleaving process generates surface potential due to the surface charges, which may develop a large effective pressure along the $b$ direction (perpendicular to the surface) and thus drive the crystal (possibly the several top layers) in the gapped state. It may be one of the possible reasons for the topological insulating behavior observed in BiPd. Such an argument is also suggested by a recent work in another noncentrosymmetric system \cite{YL Chen}. 
We have done calculations with varying lattice constant $b$, keeping total volume constant. New calculations show that strain can open up the bulk band gap.

Here we used a full-potential linearized augmented plane wave method as implemented in the WIEN2k code \cite{6}.  The generalized gradient approximation \cite{7} was also used for the exchange-correlation functional.  The spin-orbit coupling was included in a second variational way. The muffin-tin radius 2.08a$_0$ (a$_0$ being the Bohr radius), and 2.35a$_0$ and for Bi and Pd, respectively, and a plane wave cutoff RK$_{max}$ = 8 were taken in calculations that included 11 $\times$ 7 $\times$ 11 k-points. 
We note that the result for BiPd without strain from full-potential electronic structure calculations (left panel in Fig. S7) and that from the pseudopotential electronic structure calculations (right panel in Fig. S5) are quite consistent. 
In the case of compressive strain along the $b$ direction, there is a clear tendency to open up a gap (see right panel in Fig. S7). In other word, the gap region is increased due to the strain applied on BiPd system.




\vspace{0.5cm}

\subsection{\large {Supplementary References}}

\vspace{0.4cm}
Correspondence and requests for materials should be addressed to M.N. (Email:
Madhab.Neupane@ucf.edu).

\end{document}